# Possible multiband superconductivity in the quaternary carbide YRe$_2$SiC

L. R. de Faria[1], P P. Ferreira[1], L. E. Correa[1], L. T. F. Eleno[1], M. S. Torikachvili[2] and A. J. S. Machado[1]

[1] Universidade de São Paulo, Escola de Engenharia de Lorena, P. O. Box 116, Lorena, SP, Brazil

[2] Department of Physics, San Diego State University, San Diego, California 92182, USA

**Abstract**

We report for the first time the occurrence of superconductivity in the quaternary silicide carbide YRe$_2$SiC with $T_c \approx 5.9$ K. The emergence of superconductivity was confirmed by means of magnetic susceptibility, electrical resistivity, and heat capacity measurements. The presence of a well developed heat capacity feature at $T_c$ confirms that superconductivity is a bulk phenomenon, while a second feature in the heat capacity near 0.5 $T_c$ combined with the unusual temperature dependence of the upper critical field $H_{c2}(T)$ indicate the presence of a multiband superconducting state. Additionally, the linear dependence of the lower critical field $H_{c1}$ with temperature resemble the behavior found in compounds with unconventional pairing symmetry. Band structure calculations reveal YRe$_2$SiC could harbor a non-trivial topological state and that the low-energy states occupy multiple disconnected sheets at the Fermi surface, with different degrees of hybridization, nesting, and screening effects, therefore making unconventional multiband superconductivity plausible.

## 1. Introduction

The discovery of superconductivity in the RNi$_2$B$_2$C compounds (R = Y, Er, Ho, and Lu) with critical temperatures ($T_c$) as high as 15 K served as a compelling motivating factor for studies in related quaternary materials [1,2]. The crystal structure is tetragonal and it can be described as a variant of the ThCr$_2$Si$_2$-type (space group SG I4/mmm), consisting of planar layers of RC stacked between layers of Ni$_2$B$_2$ with nickel ions tetrahedrally coordinated to four boron atoms [3]. The interest in borocarbides is twofold. First, the $T_c$ values are moderately high for intermetallic compounds. And secondly, coexistence of superconductivity and magnetic ordering was observed for some members of this family, when the R ions carry a magnetic moment, e.g., R = Tm, Er, Ho, and Dy [1,4].

Other superconducting quaternary borocarbides with different transition metals replacing nickel (e.g. Rh, Pd, Pt, Re, and Ir) were reported, [5–8] as well as some boronitrides, in which N replaces C [9]. While superconductivity has been observed in quaternary compounds with the replacement of the metallic elements or carbon, studies with partial or total substitution of boron are limited, though filamentary superconductivity was reported for YNi$_2$Si$_2$C and YCu$_2$Si$_2$C [10].

YRe$_2$SiC is a quaternary silicide carbide with orthorhombic structure, as reported by Hüfken et al. [11]. This structure has DyFe$_2$SiC (SG Cmcm) as prototype [12] and it can be described as a filled variant of the Re$_3$B structure, where Re$_2$Si units form a quasi-two-dimensional layer with a puckered hexagonal arrangement [13]. These sheets are connected by carbon ions through Re-C-Re bridges, while the Y atoms fill available atomic positions in between the layers. Many other compounds crystallize in this structure, where other lanthanides, and the actinides Th and U replace Y, while Re can be replaced by other transition metals, e.g. Fe, Mn, Ru and Os. Also, phosphorus can replace silicon, making up a class of more than 80 compounds [12–15]. In spite of these new quaternary compounds providing a vast phase space for studying electronic and magnetic properties, very few were studied experimentally [14,16].

In this work we provide a detailed study of YRe$_2$SiC, which represents the first example of a bulk superconducting quaternary silicide carbide in literature. Resistivity measurements show a clear superconducting phase transition at $T_c \approx 5.9$ K. The two anomalies found in the specific heat data at $T_c$ and $\approx 0.5T_c$, together with the unusual upturn below $T_c$ of the upper critical field $H_{c2}$, suggest the presence of a multiband superconducting ground state. Additionally, the linear dependence of the lower critical field $H_{c1}$ with temperature suggest an unconventional pairing mechanism. These results are consistent with the band structure calculations that we report here for the first time, which shows that YRe$_2$SiC can harbor a non-trivial topological state, opening the way for further theoretical and experimental investigations.

## 2. Experimental Details

The YRe$_2$SiC and YRe$_{1.8}$SiC samples for this study were synthesized in an arc furnace. Stoichiometric amounts of high purity Y, Re, Si and C (> 99.9%) were placed on a water-cooled copper hearth, and melted together several times, turning over each time, in a Ti-gettered ultra-high pure (UHP) argon atmosphere. In order to gauge the effect of the cooling rate on the microstructure, a second YRe$_2$SiC sample was prepared in a modified arc furnace allowing cooling at a much lower rate. The weight losses were lower than 1%, and they were not corrected with additional constituents. All samples were sealed in quartz ampoules under 0.5 atm of UHP argon, and annealed at 1000 ºC for 72 h. A third YRe$_2$SiC sample was prepared under slow reduction of the current in the arc furnace, in order to allow an even slower cooling, followed by a sub-solidus heat treatment, which occurs when the sample's temperature is slightly lower than the solidification temperature, allowing for high diffusion rates. X-ray powder diffraction scans were taken with a Panalytical Empyrean diffractometer with either Cu-K$\alpha$ or Mo-K$\alpha$ radiation. All diffractograms were analyzed using the HighScore Plus software, based on the Rietveld methodology for structural refinement. Given that the melting temperature of YRe$_2$SiC is fairly high, the cool down process is fast and inhomogeneous across the different parts of the ingots, and a microstructural analysis was in order. The ingots were cut in half, mounted and polished for a scanning electron microscope (SEM) analysis, which was performed with a Hitachi TM-3000 SEM, equipped with an Oxford energy dispersive spectrometer (EDS).

Magnetization ($M$) and electrical resistivity ($\rho$) measurements were carried out in thin and elongated sections ($\approx$ 1.0 x 0.5 x 0.3 mm$^3$) using a Physical Property Measurement System (PPMS) from Quantum Design. Zero-field-cooled (zfc) and field-cooled (fc) magnetization as a function of temperature ($T$) in 20 Oe, and magnetization as a function of the applied field ($H$) data were collected using the vibrating sample magnetometer (VSM) option. The $\rho(T)$ data was collected using the standard four-probe method between 2.0 and 300 K. The temperature dependence of the lower critical field ($H_{c1}$) was extracted from the magnetization curves below $T_c$, and the upper critical field $H_{c2}$ vs $T$ data were extracted from the midpoint of the $\rho(T)$ transitions to the $\rho$ = 0 state. Heat capacity ($C_p$) measurements in magnetic field were carried out with the calorimeter option of the PPMS, which uses a relaxation method. $C_p$ measurements near $T_c$ were used to confirm that superconductivity in YRe$_2$SiC is a bulk phenomenon, to reveal a second feature below $T_c$ suggestive of multiband behavior, and to deconvolute the electronic and phonon contributions.

## 3. Experimental results and discussion

All samples for this study have the desired YRe$_2$SiC phase (1211, orthorhombic, SG Cmcm), though a small amount of a Re-rich solid-solution (Re$_{ss}$) phase can be identified in X-ray diffraction scans and SEM micrographs. The microstructural analysis suggests that the 1211 phase forms from an incongruent phase transition upon solidification. The dendritic morphology of the Re$_{ss}$ impurity phase suggests that it is the first to solidify from the melt. Given the very high melting point and the complex convolution of possible phases, it is quite challenging to obtain single-phase materials. Detailed X-ray diffraction scans, microstructural data, as well as $\chi(T)$ and $\rho(T)$ measurement for the samples of this study are in the Supplementary Materials, and these data show convincingly that YRe$_2$SiC is superconducting, though the superconducting properties are quite sensitive to phase purity. Rapid cooling in both, stoichiometric and Re-deficient YRe$_{1.8}$SiC samples, seem to favor a fine morphology dense with microstructural defects, leading to weakened superconducting properties. In contrast, slow cooled samples have optimized superconducting properties, in spite of being more amenable to impurity phases. The superconducting volumes are higher, and the resistivity transitions are narrower in the slow cooled samples, in spite of the extra phases. The SEM micrographs suggest that the YRe$_2$SiC phase in the slower cooled samples are larger and have better continuity. Therefore, we focused our study of the superconducting properties on the slower cooled samples. The temperature dependence of the electrical resistivity for the slow-cooled YRe$_2$SiC between 2 and 300 K is shown in figure 1(a). $\rho(T)$ displays a slight downward curvature in the 50 – 300 K temperature range. This type of behavior in a intermetallic compound is consistent with presence of extra phases, voids, and a certain level of disorder. A superconducting transition can be clearly seen with onset at $T_c \approx 5.9$ K, and the $\rho = 0$ value is reached at $\approx 4.7$ K. The normalized values of $\rho/\rho_{15K}$ near $T_c$ for magnetic fields $0 \leq \mu_0 H(T) \leq 5.0$ T are shown in figure 1(b).

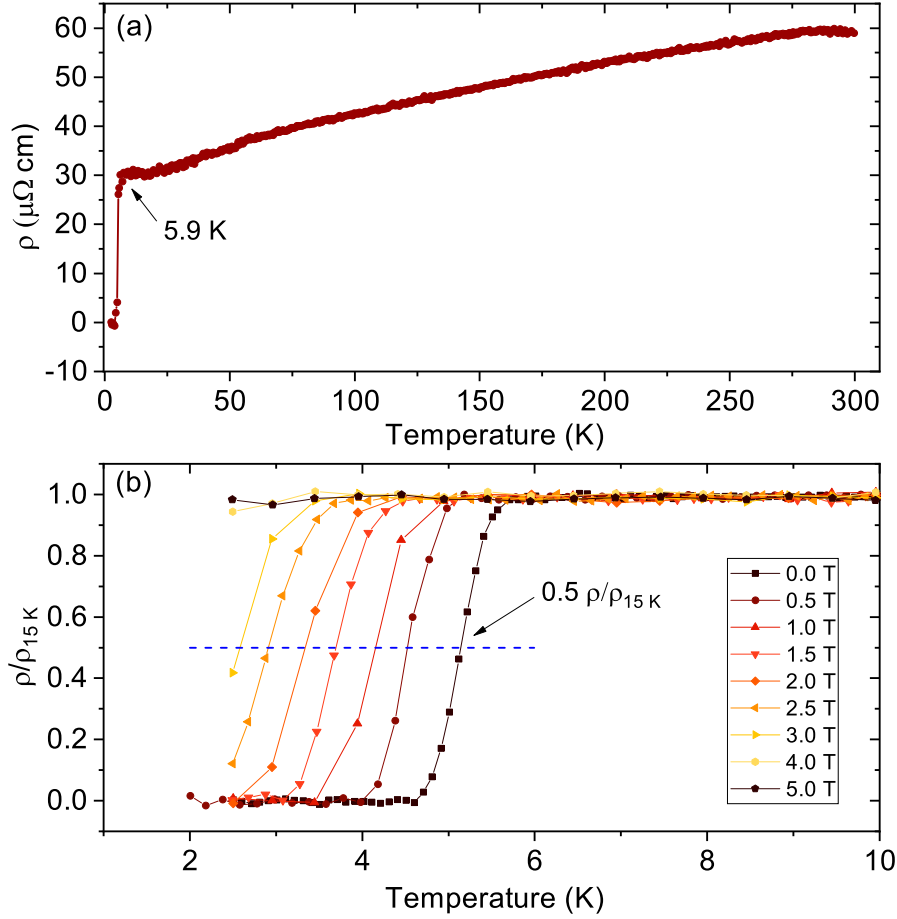

Figure 1. (a) Temperature dependence of the electrical resistivity for the slow-cooled YRe$_2$SiC between 2 and 300 K; (b) normalized resistivity $\rho/\rho_{15K}$ versus temperature in fields up to 5 T.

Extracting $T_c$ at the various fields of figure 1(b) from the midpoint of the superconducting transitions, and from the peak features of the heat capacity, the $H_{c2}$ vs $T/T_c$ phase diagram separating the normal and superconducting phases can be constructed as shown in figure 2.

A fit of the $H_{c2}$ data to the equation $H_{c2}(T)=H_{c2}(0)[1-(T/T_c)^2]$ which is consistent with the Ginzburg-Landau equations for a single-band isotropic superconductor, results in the red dashed line of figure 2. The value $H_{c2}(0) = 2.5$ T for this fit was taken from the Werthamer, Helfand and Hohenberg (WHH) model using equation 1 below [17], which is applicable in the dirty-limit

$$H_{c2}(0) = -0.693 T_c [dH_{c2}/dT]_{T=T_c} \quad (1)$$

This fit obviously underestimates the values of $H_{c2}$, and it doesn't reflect the positive upturn that becomes more noticeable at lower temperatures.

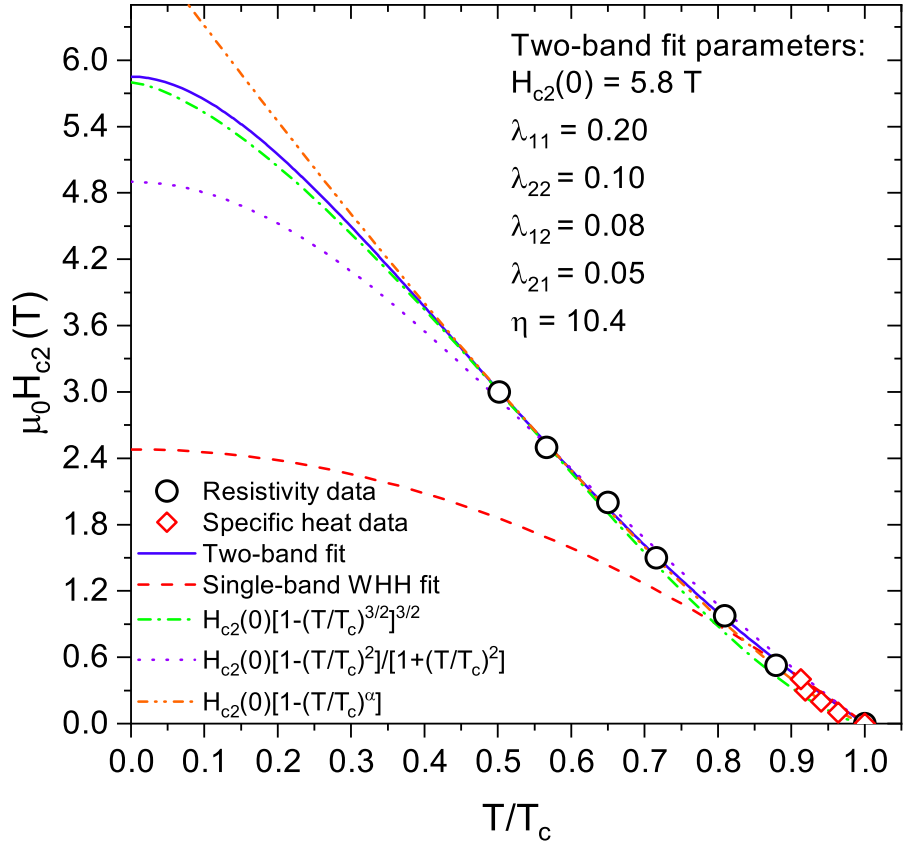

Figure 2. Upper critical field $H_{c2}$ vs. $T/T_c$ for YRe$_2$SiC. The red-dashed line is a fit to the experimental data using a single-band approach, while the blue-solid line corresponds to the two-band fit of equation 5 with parameters $\lambda_{11}$=0.20, $\lambda_{22}$=0.10, $\lambda_{12}$=0.08, $\lambda_{21}$=0.05 and $\eta$=10.4. Empirical fits related to equations 2–4 are also exhibited.

The positive curvature of $H_{c2}(T)$ has been addressed in theoretical and empirical models consistent with a two-band scenario, and the Ginzburg-Landau (GL) theory [18–20]. For instance [18],

$$H_{c2}(T)=H_{c2}(0)[1-(T/T_c)^2]/[1+(T/T_c)^2] \qquad (2)$$

is obtained from the GL theory in the presence of two superconducting order parameters. The expression

$$H_{c2}(T)= H_{c2}(0)(1-(T/T_c)^{3/2})^{3/2} \qquad (3)$$

is applicable for unusual pairing mechanisms [19], and

$$H_{c2}(T)=H_{c2}(0)(1-T/T_c)^\alpha \qquad (4)$$

is an empirical expression consistent with the solution of the linearized Eliashberg equations for $T > 0.3T_c$ [20]. Equations 2–4 all yield relatively good fits to the $H_{c2}$ data for $T/T_c > 0.5$, as can be seen in figure 2.

The positive curvature of $H_{c2}(T)$ for YNi$_2$B$_2$C and LuNi$_2$B$_2$C has been treated in terms of an effective two band model that can be solved through the linearized Eliashberg equations [21,22], although such approach requires several experimental inputs, such as the spectral function $\alpha^2 F(\omega)$. Unfortunately, the latter is not available for YRe$_2$SiC yet. However, linearized Eliashberg equations in the dirty limit reduce to the Usadel equations [23], and the temperature dependence of $H_{c2}$ may be described by

$$a_0[\ln t + U(h)][\ln t + U(\eta h)] + a_1[\ln t + U(h)] + a_2[\ln t + U(\eta h)] = 0 \qquad (5)$$

where $h = HD_1/2\phi_0 T$, $\eta = D_2/D_1$, $a_1 = 1 + \lambda_-/\lambda_0$, $a_2 = 1 - \lambda_-/\lambda_0$, $a_0 = 2w/\lambda_0$, $t = T/T_c$ and $U(x) = \psi(x + 1/2) - \psi(1/2)$. In $a_{0,1,2}$ we have $\lambda_\pm = \lambda_{11} \pm \lambda_{22}$, $w = \lambda_{11}\lambda_{22} - \lambda_{12}\lambda_{21}$, and $\lambda_0 = (\lambda_-^2 + 4\lambda_{12}\lambda_{21})^{1/2}$, while $\psi(x)$ is the digamma function, $\phi_0$ is the magnetic flux quantum. Finally, $\lambda_{ii}$ and $\lambda_{ij}$ ($i \neq j$) are the intraband and interband electron coupling constants and $D_1$ and $D_2$ are the electron diffusivities in bands 1 and 2, respectively.

A two-band fit of $H_{c2}(T)$ for YRe$_2$SiC using equation 5 is in excellent agreement with the data, and it replicates the positive curvature, as shown in figure 2. The fitting parameters are $H_{c2}(0) = 5.8$ T, $\lambda_{11} = 0.20$, $\lambda_{22} = 0.10$, $\lambda_{12} = 0.08$, $\lambda_{21} = 0.05$, with $\eta = 10.4$. This model has been applied successfully in known multiband materials such as MgB$_2$ and iron-based superconductors [23–26]. The ratio $a_2/a_1 = 0.234$ in YRe$_2$SiC is larger than the ratio $D_1/D_2 = 1/\eta = 0.096$, which is an indication that $H_{c2}(T)$ near $T_c$ is governed by the smaller gap, which has higher diffusivity [27]. The values of $\lambda_{12}$ and $\lambda_{21}$ are indicative of a weak interband scattering, which, in fact, favors the emergence of two superconducting gaps at the Fermi surface. This two-band model is particularly powerful, since it provides a large amount of information on the materials behavior from a relatively simple measurement, given that the values become constrained to several relations that give physical significance to the model [23]. Further analysis of the angle dependence of the $H_{c2}(T)$ would be useful. However, it is currently precluded given the challenge of growing single crystals. It is important to point out that a positive curvature may arise from other reasons, such as paramagnetic effects, strong coupling or an anisotropic superconducting gap [28–30].

Magnetization data for YRe$_2$SiC are shown in figure 3. The temperature dependence of the magnetic susceptibility in figure 3(a) shows a clear onset of superconductivity near $T_c \approx 5.4$ K. The zfc data suggest a magnetic shielding fraction of about 95 % at 2.5 K in a 20 Oe magnetic field. The inset shows a 3.0 K isotherm of $M$ vs $H$, where the hysteresis loop for the Meissner state is well developed, indicating type-II superconductivity. A complete set of isothermal $M$ vs $H$ curves in the $T$- range from 2 to 5 K is shown in figure 3(b).

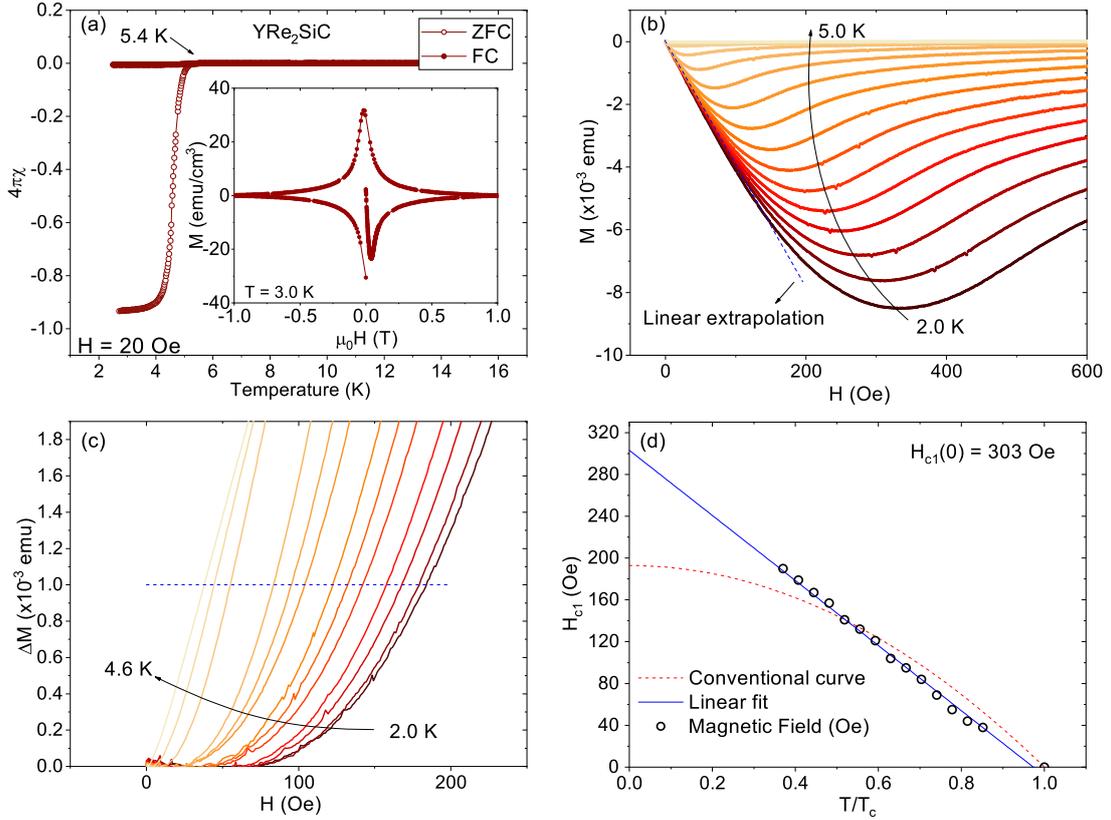

Figure 3. (a) ZFC and FC magnetic susceptibility for polycrystalline YRe$_2$SiC in a 20 Oe magnetic field. The two curves split at $T_c \approx 5.4$ K, and the shielding fraction at 2.5 K is about 95%. Inset: isothermal $M$ vs $H$ loop at 3.0 K. (b) $M$ vs $H$ isotherms in the 2 – 5 K temperature range for slow-cooled YRe$_2$SiC; (c) $\Delta M$ vs $H$ plots for different temperatures, where $\Delta M$ is the difference between the experimental data and the extrapolated values from the low field $M$ vs $H$ Meissner lines. The $H_{c1}$ values at each temperature were taken from the fields where $\Delta M = 10^{-3}$ emu; (d) $H_{c1}$ vs $T/T_c$ obtained from (c). The blue corresponds to a linear fit of the data, while the red line corresponds to the conventional fit.

In order to determine the temperature dependence of $H_{c1}$, we extracted the $H_{c1}$ values at each temperature from the $M(H)$ isotherms of figure 3(b), by taking the values where the data first deviates by $\Delta M = 10^{-3}$ emu from the linear behavior at low field [31]. This criterion is indicated by the dashed horizontal line in figure 3(c), which shows isotherms of $\Delta M$ vs $H$.

A plot of $H_{c1}$ vs $T/T_c$ for YRe$_2$SiC is shown in figure 3(d), where the red dashed line corresponds to an isotropic single-band fit ($H_{c1}(T)=H_{c1}(0)[1-(T/T_c)^2]$). This is obviously a bad fit, as its parabolic behavior is in contrast with the linear behavior of $H_{c1}$ vs $T/T_c$. A similar linear behavior of $H_{c1}$ vs $T/T_c$ is observed in YNi$_2$B$_2$C [32] and MgB$_2$, where it may be due to the existence of either an anisotropic s-wave gap function or an unconventional gap symmetry which may result in a linear dependence of superfluid density with temperature [31], though further work is needed to confirm the nature of what is observed in YRe$_2$SiC. By carrying out a linear extrapolation of the $H_{c1}$ vs $T/T_c$ data to low temperatures, $H_{c1}(0)$ can be estimated to be $\approx$ 303 Oe.

The $H_{c2}(0)$ value of 5.8 T obtained from equation 5 can be used to estimate the Ginzburg-Landau coherence length from equation 6 below

$$H_{c2}(0)=\frac{\phi_0}{2\pi\xi_{GL}^2} \tag{6}$$

where $\phi=h/2e$ is the quantum flux, yielding $\xi_{GL}$ = 74 Å. This is close to $\approx$ 97 Å value reported for YNi$_2$B$_2$C [33]. Considering that $H_{c1}(0) \approx$ 303 Oe, the values of the penetration depth $\lambda_{GL}$ can be estimated from

$$H_{c1}(0)=\frac{\phi_0}{4\pi\lambda_L^2}\ln\frac{\lambda_{GL}}{\xi_{GL}} \tag{7}$$

yielding $\lambda_{GL}$ = 1212 Å, which is also close to the observed value of 1200 Å for YNi$_2$B$_2$C [33]. The value of $\kappa_{GL} = \lambda_{GL}/\xi_{GL}$ was calculated to be $\kappa_{GL}$ = 16 for YRe$_2$SiC, which is consistent with type-II superconductivity. The thermodynamic critical field can be calculated from equation 8

$$H_{c1}H_{c2}=H_c^2\ln\kappa_{GL} \tag{8}$$

yielding $H_c$ = 252 mT.

In order to ascertain that the superconductivity in YRe$_2$SiC is a bulk phenomenon, we carried out measurements of heat capacity ($C_p$). The plot of $C_p/T$ vs $T^2$ in $H$ = 0 displayed in figure 4(a) clearly shows two anomalies consistent with superconductivity, i.e., a small feature centered near 2.8 K, and a much larger centered near 4.5 K. These anomalies are suppressed by $H = 5$ T, a field that exceeds $H_{c2}$ for $T > 1.3$ K. The magnitude of the jump corresponding to the larger anomaly in $C_p$ is $\Delta C_e/\gamma T \approx 0.8$ (figure 4(b)), about 55% of the expected value of 1.43 for the BCS theory. It is plausible that this lower value may be associated with a non-superconducting fraction of material, as frequently observed in other materials, e.g $HfV_2Ga_4$, $LaOs_4As_{12}$, and $Nb_2PdS_5$ [34–36].

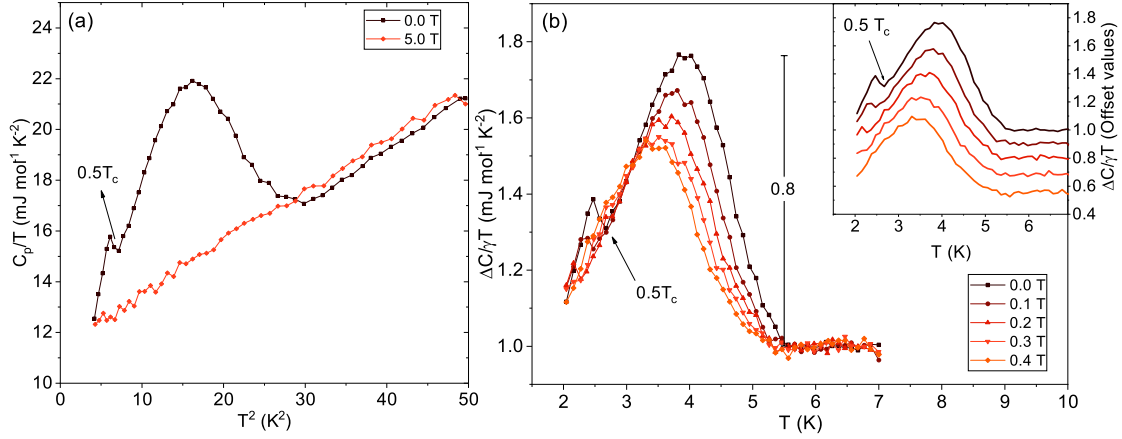

Figure 4. (a) Heat capacity $C_p/T$ vs $T^2$ for $YRe_2SiC$ in $H = 0$ and 5 T; and (b) $\Delta C_e/\gamma T$ vs $T$ in $H = 0 - 0.4$ T. The data in the inset are offset for clarity. The superconducting transition and the anomaly centered near 2.8 K are both suppressed by the magnetic field.

The appearance of the second feature in the $C_p$ data near $0.5T_c$ suggests that the superconductivity in $YRe_2SiC$ may be considered in the context of a multiband effect. As shown in the inset of figure 4(b), both $C_p$ anomalies are driven to lower temperatures by the magnetic field, consistently with what should be expected if they are both due to superconductivity.

At first, there is no evidence of a second bulk superconducting phase neither from XRD nor EDS which could manifest itself on heat capacity (See supplementary material for details). Other effects such as nematic superconductivity (or other topological phase transitions) or Schottky anomalies also seem unlikely, since those types of transition tend to be either insensitive or to shift towards higher temperatures in the presence of magnetic field [37,38]. Thus, it seems that the second peak originates from $YRe_2SiC$.

Field-dependent secondary anomalies that dislocate towards lower temperatures are commonly seen in multiband superconductors such as $MgB_2$ [39], $HfV_2Ga_4$ [34], $LaOs_4As_{12}$ [35], and $Nb_2PdS_5$ [36] and are attributed to the opening of a secondary small gap that surfaces in the $C_p$ behavior at low temperatures, while the larger gap dominates the behavior at higher temperatures. In the intermediate temperatures, a crossover of both contributions takes place, in which interband coupling and impurity scattering smoothes out the transitions, smearing the lower-$T$ anomaly, such that instead of a secondary peak one observes a kink [40]. In this scenario, the observation of a small peak rather than a shoulder in the heat capacity of YRe$_2$SiC is consistent with the small values of interband coupling parameters $\lambda_{12} = 0.08$ and $\lambda_{21} = 0.05$ calculated from the $H_{c2}(T)$ data.

The value of the Sommerfeld coefficient $\gamma$ can be extracted from an extrapolation of the normal phase $C_p/T$ vs $T^2$ to $T = 0$, yielding 10.4 mJ/mol.K$^2$. The phonon contribution can be extracted from the slope, yielding $\beta = 0.233$ mJ/mol.K$^4$, from which the value of the Debye temperature $\Theta_D$ can be estimated to be ≈ 347 K.

To the best of our knowledge, this is the first time superconductivity is reported in YRe$_2$SiC. A summary of the superconducting parameters is given in Table 1.

Table 1. Observed superconducting properties for YRe$_2$SiC.

| Parameter | YRe$_2$SiC |
|---|---|
| $T_c$ | 5.9 K |
| $\mu_0 H_{c1}(0)$ | 30.3 mT |
| $\mu_0 H_{c2}(0)$ | 5.8 T |
| $\mu_0 H_c(0)$ | 252 mT |
| $\lambda_{GL}$ | 1212 Å |
| $\xi_{GL}$ | 74 Å |
| $\kappa_{GL}$ | 16 |
| $\lambda_{11}$ | 0.20 |
| $\lambda_{22}$ | 0.10 |
| $\lambda_{12}$ | 0.08 |
| $\lambda_{21}$ | 0.05 |
| $\eta$ | 10.4 |
| $\gamma$ | 10.4 mJ/mol.K$^2$ |
| $\beta$ | 0.233 mJ/mol.K$^4$ |
| $\Theta_D$ | 347 K |

## 4. Electronic structure calculation

In order to probe the consistency of the experimental findings, we carried out fully-converged first principles electronic structure calculations within the scope of the Density Functional Theory (DFT) [41,42] using scalar relativistic ultrasoft pseudopotentials [43], with the parametrization of Perdew-Zunger for the exchange and correlation effects [44], as implemented in the Quantum Espresso code [45,46].

The non-spin-polarized electronic band structure along the high-symmetry points in the first Brillouin zone and the projected electronic density of states (PDOS) for YRe$_2$SiC are shown in figure 5(a). There are four distinct bands crossing the Fermi level, originating multiple disconnected electron- and hole-pockets. The total density of states at the Fermi energy ($E_F$) is 8.4 states/eV, which translates into a Sommerfeld coefficient value $\approx$ 19.8 mJ/mol.K$^2$ using the free electron model approximation, which is within a factor of 2 of the 10.4 mJ/mol.K$^2$ determined experimentally. The discrepancy between the Sommerfeld coefficient obtained with DFT calculations and heat capacity is usually related to the moderate mass enhancement of the electron-phonon coupling parameter and spin-orbit coupling effects [47].

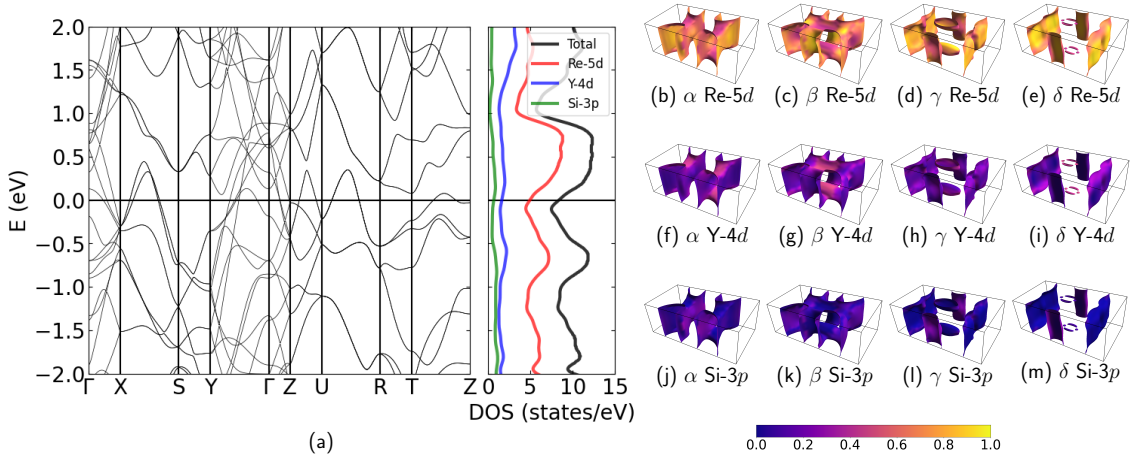

**Figure 6.** (a) Non spin-polarized electronic band structure and projected density of states for YRe$_2$SiC; and (b) projected Fermi surface sheets ($\alpha$, $\beta$, $\gamma$ and $\delta$) on Re-5$d$ orbitals (b)-(e), Y-4$d$ orbitals (f)-(i) and Si-3$p$ orbitals (j)-(m).

The Re-5$d$ manifold corresponds to 59.0% of the carriers, while 19.7 and 8.3% of the low-energy states are derived from Y-4$d$ and Si-3$p$, respectively. The Fermi surface projected onto the electronic character is shown in figures 5(b)-(m). Given that YRe$_2$SiC has electrons occupying multiple disconnected sheets in the Fermi surface, with different degrees of hybridization, screening effects and nesting vectors, multiband superconductivity is plausible and consistent with the behaviors of $H_{c1}$ and $H_{c2}$ [48–53]. In fact, the small contributions of Y-4$d$ and Si-3$p$ wavefunctions in the Fermi surface is consistent with the smaller second anomaly found in heat capacity measurements [34, 53]. The topography of the Fermi surface also favors nesting-induced instabilities in both spin and charge susceptibilities, possibly leading the system on the verge of unconventional superconductivity [52–55], as suggested by the linear dependence of $H_{c1}$ with temperature.

Additionally, the non-spin-polarized electronic band structure harbors several features suggestive of a non-trivial metallic phase, such as the linear band-crossings around $E_F$, four-fold nodal lines band-degeneracies along X-S-Y, Z-U and R-T-Z, and eight-fold nodal lines along U-R, including spin-degeneracy, as shown in figure 5(a). [54–56]. Therefore, the spin-orbit coupling can give origin to topological band inversions due to the lack of non-symmorphic symmetries [57–62], opening the way for further theoretical and experimental investigations in YRe$_2$SiC beyond conventional superconductivity.

## 5. Conclusions

We carried out a systematic study of the synthesis and superconducting properties of the orthorhombic quaternary carbide YRe$_2$SiC. This 1211 phase has a $T_{c,\text{onset}} \approx 5.9$ K, $H_{c2}(0) \approx 5.8$ T, and $\xi_{GL} \approx 74$ Å. Given the very high melting point and the quaternary nature, single phase samples are difficult to obtain by arc melting. Small amounts of Re$_{ss}$ (mostly a solid solution of $\approx 4$ at% Y in Re), and other minority phases, can be minimized by appropriate processing, but not completely eliminated. The YRe$_2$SiC samples with optimized superconducting properties were obtained in a modified arc furnace where a reduced cooling rate was possible. In spite of a small amount of unidentified phases, larger areas of the 1211 phase were identified near the center of the ingots. These sections showed almost full diamagnetic screening below $T_c$, and narrow resistivity transitions, with $\rho = 0$ state being reached near 4.8 K. The normal state resistivity drops slightly with temperature. It is difficult to ascertain at this point whether the excessive scattering is dominated by impurities and defects, or perhaps that this is a strongly correlated electron system. $H_{c2}(T)$ deviates from the most typical behavior in conventional superconductors; it shows a positive curvature below $T_c$ reminiscent of other exotic superconductors. Low field magnetization curves reveal a linear dependence of $H_{c1}$ with temperature. This behavior is not expected for conventional $s$-wave superconductors, suggesting possibly that the pairing may be unconventional. The presence of small secondary peak in $C_p$ suggests the opening of a second superconducting gap near $0.5T_c$, and in turn that YRe$_2$SiC is a multiband superconductor. However, in light of the small amount of impurity phases that maybe present, a phase transition in one of them cannot be completely ruled out as the source of this feature. Band structure calculations reveal four distinct band crossings at the Fermi level, giving credence to the possibility of multiband superconductivity and non-trivial topological effects. Albeit challenging due to the complexity of the quaternary phase diagram, an effort to obtain single phase or single crystalline YRe$_2$SiC is in order, given the interest in addressing the issues of multiband superconductivity, and unconventional pairing mechanisms.

## Acknowledgements

We gratefully acknowledge the support for this research from the São Paulo Research Foundation (FAPESP, Brazil), under grants No. 2016/11774-5, 2019/14359-7, 2019/17878-5, 2019/05005-7, and 2020/08258-0, and the Brazilian National Council for Scientific and Technological Development (CNPq) under grants No 431868/2018-2, 302149/2017-1 and 132934/20191, and the high-performance computing resources from STI, Universidade de São Paulo, Brazil. MST is grateful for the hospitality during his visit to the Escola de Engenharia de Lorena, Universidade de São Paulo, Brazil.